\documentclass[12pt]{article}
\usepackage{epsfig}

\begin{document}
\title{A numerical study of the Schr\"odinger-Newton equation \newline
 2: the time-dependent problem}
\author{R Harrison, I Moroz and K P Tod \\ Mathematical Institute \\ St Giles \\ Oxford OX1
3LB \\\ }
%{\em tel 01865-273525} \\ {\em fax 01865-273583}\\ 
%{\em e-mail tod@maths.ox.ac.uk}}

%\date{}

\maketitle

\begin{abstract}

We present a numerical study of the time-dependent SN
equations in three dimensions with three kinds of symmetry: 
spherically symmetric, axially symmetric and
translationally symmetric. We find that the
solutions show a balance between the dispersive tendencies of the
Sch\"odinger equation and the gravitional attraction from the Poisson
equation. Only the ground state is stable, and lumps of probability
attract each other gravitationally before dispersing.

\end{abstract}

\section{Introduction}
Ths is the second in a series of papers presenting a numerical study
of the Schr\"odinger-Newton (or SN) equations. In the first
(\cite{hmt1}) we reviewed known results on the SN equations and 
then analysed the linear stability of the
spherically-symmetric stationary states. We found that the ground
state is linearly stable, in  that all the eigenvalues for linear
perturbations were purely imaginary, while all the  higher states are
unstable. The $(n+1)$-th state, or equivalently the $n$-th excited
state, has $n$ quadruples $(\lambda, -\lambda, \bar{\lambda},
-\bar{\lambda})$ of complex eigenvalues with nonzero real
part. We now want to go further in two directions. On the one hand we shall
consider the full nonlinear evolution of the SN equations and on the
other we shall move away from spherical symmetry, allowing for two space
variables. 

In the evolution of spherically-symmetric data we shall find that
the picture from linear theory is confirmed: the ground state is
stable but slight perturbations of the higher states decay, with
probability density escaping off the grid to leave a `nugget'
consisting of the ground state rescaled to have total probability less
than one and situated at the origin. For more general
spherically-symmetric data the picture is confirmed: solutions 
typically disperse leaving a nugget of rescaled 
ground state. (By a rescaled ground
state we mean the following: if $\psi_0(r)$ is the spatial part 
of the wave function for the
ground state, normalised to have total probability equal to 
one and with conserved energy ${\mathcal{E}}_0$
in the terminology of \cite{hmt1}, then $p^2\psi_0(pr)$
with $p$ a positive constant, is a stationary solution with 
conserved energy $p^3{\mathcal{E}}_0$
and total probability $p$.)

We can introduce a new space variable in two ways. We may consider
axially-symmetric solutions in 3-dimensions. In this case there are
new 3-dimensional stationary states, somewhat analogous to the 
axially-symmetric
solutions of the hydrogen atom, and with nonzero expectation for the angular
momentum. These turn out to be unstable under evolution, as one would
expect. For the solution most analogous to a pure dipole, the
evolution clearly shows the two `lumps' of probability falling from
rest into
each other and evolving towards the ground state, with some scatter.

Alternatively we may consider translation-invariant solutions in
3-dimensions or equivalently the SN equations in $2+1$-dimensions. (Of
course the solutions are not normalisable as $3+1$-dimensional
solutions but this is still an interesting problem.) In this case we
can find rigidly-rotating stationary states like two lumps of
probability rotating around each other. When these are evolved with
the time-dependent SN equations they
prove to be unstable, so that it is
possible to arrange for two lumps of probability to be in orbit
around each other at least for a while before they merge.\\

In summary, the picture that we find is of a system with dispersive
tendencies because of the Schr\"odinger equation and attractive or
concentrating tendencies from the gravitational attraction. There
appear to be infinitely many stationary states, all of them unstable
except for the ground state. General data evolve to leave some
residual probability in a rescaled ground state with the rest of the
probability dispersing. Lumps of probability in the initial data can
attract each other and even orbit each other but eventually the
dispersive tendencies win.\\

The plan of the paper is as follows. We shall end this
Introduction with an analytic result bounding the residual 
probability left on the
grid at late times. In Section 2 we introduce the numerical method
which we shall use to evolve the spherically-symmetric SN
equations. The results of this evolution are presented in Section
3. In Section 4 we present the results of numerically evolving
3-dimensional axisymmetric solutions and in Section 5 we present the
results of numerically evolving the 2-dimensional SN equations.

For the analytic calculation then, suppose
that the picture presented above holds in general so that 
arbitrary initial data evolve to give a scattering solution which
disperses to infinity and leaves a rescaled
ground state. Then we can obtain a bound on the probability remaining
in the ground state provided the initial energy is negative.
For suppose the initial value of the conserved energy is
${\mathcal{E}_I}$ then since this is conserved, we can calculate at
late times to find
\[\mathcal{E}_I = \mathcal{E}_S + p^3\mathcal{E}_0\]
where ${\mathcal{E}_S}$ is the energy in the scattering solution,
which we suppose to be positive, ${\mathcal{E}_0}$ is the (negative)
energy of the ground state and $p$ is the probability left in the
ground state. (The ground state is strongly peaked at the origin so
that the cross-term in $\mathcal{E}_I$ is zero.) Taking 
account of all the signs, if  $\mathcal{E}_I<0$  we can rearrange this
to read
\begin{equation}
p^3 > \frac{|\mathcal{E}_I|}{|\mathcal{E}_0|} \label{bound}
\end{equation}
which is the desired bound.

\section{Numerical methods for the spherically-symmetric evolution}

Recall from \cite{hmt1} that the SN equations are the system
\begin{eqnarray}
i\frac{\partial\psi}{\partial{}t} &=& -\nabla^2\psi +\phi\psi,\label{ndimS}\\
\nabla^2\phi &=& |\psi|^2. \nonumber
\end{eqnarray}
In this section and the next, we are concerned only with the
spherically-symmetric case and in this case (\ref{ndimS}) can be
simplified as
\begin{eqnarray}
i\frac{\partial{}u}{\partial{}t} &=&
-\frac{\partial{}^2u}{\partial{}r^2} + \phi{}u,\label{ss1}\\
\frac{\partial^2(r\phi)}{\partial r^2} &=&  \frac{|u|^2}{r},\label{ss2}
\end{eqnarray}
where $u=r\psi$.

To solve this system, we shall use a Crank-Nicholson method for the 
time-evolution of the
Schr\"odinger equation and the spectral method of \cite{hmt1} for the
Poisson equation. Schematically, this takes the form
\begin{eqnarray}
2i\frac{u^{n+1} - u^{n}}{\delta{}t} &=& -D^2(u^{n+1})-D^2(u^{n})
\label{cn1}\nonumber \\ &&+\phi^{n+1}u^{n+1} + \phi^n{}u^n,  
\end{eqnarray}
where, with the spectral method of \cite{hmt1}, 
$D^2$ is the second-derivative matrix for Chebyshev polynomials. The 
Crank-Nicholson method is second-order accurate in the
time and preserves the normalisation of the wave function.  

Now
we need an iterative method to find $\phi^{n+1}$. We write
$\phi^{n+1}_k$ and $u^{n+1}_k$ for the iterates and make the
initial choice $\phi^{n+1}_0 = \phi^{n}$, then use the spectral method
to solve (\ref{cn1}) for
$u^{n+1}_0$, use this in (\ref{ss2}) to improve
$\phi^{n+1}_0$ to $\phi^{n+1}_1$, and solve (\ref{cn1}) to find 
$u^{n+1}_1$. This
cycle can be repeated until the desired degree of convergence is
reached.

The boundary conditions are that $u$ and $r\phi$ should 
vanish at the origin and the outer edge $r=L$ of the grid. We 
expect there to be an outgoing flux of probability and we
need to prevent it from reflecting back off of the boundary. For 
this we use the `sponges' of Bernstein et al \cite{mp98}. The idea is 
to replace the $i$ on the left-hand-side in (\ref{ss1}) by
$i+s(r)$ where $s(r)$ is a positive function large (of order one) near the
boundary but small close in. We typically use $s(r)=ae^{b(r-L)}$ for
constants $a$ and $b$. 
By this means the Schr\"odinger equation is
converted to a heat equation near the boundary and the outgoing flux
is absorbed.\\

We can perform various checks on the method outlined here. 
To check that the sponges are working as desired, we temporarily 
set $\phi~=~0$ and consider the numerical
evolution corresponding to the following explicit spherically-symmetric 
solution of the
zero-potential Schr\"odinger equation:
\begin{eqnarray}
r\psi = u = \frac{C\sqrt{\sigma}}{(\sigma^2 +2it)^\frac{1}{2}}
[\exp(-\frac{(r-vt-a)^2}{2(\sigma^2 +2i{}t)} +\frac{i{}vr}{2} -
\frac{i{}v^2t}{4}) \nonumber \\ 
- \exp(-\frac{(r+vt+a)^2}{2(\sigma^2+2i{}t)} -
\frac{i{}vr}{2} - \frac{i{}v^2t}{4})] 
\label{lexp}
\end{eqnarray}     
where $C$ is a normalisation constant (which can be found
explicitly). 
This solution is a
spherically-symmetric bump with a Gaussian profile 
centred initially at $r=a$ and moving
radially with
velocity $v$. It both disperses and moves off the
grid. Computing it is a test for the Schr\"odinger evolution, and having it
move smoothly off the grid is a test for the sponges. The initial
data for the solution (\ref{lexp}) will be evolved in the full
evolution in Section 3.

Next, we can include a fixed $\phi$ in (\ref{ss1}), dropping (\ref{ss2}) and
jointly 
test the Schr\"odinger evolution and the sponges. Finally, we 
can test convergence of the method by varying the time step and/or
the number of Chebyshev points chosen in the spectral method.

\section{Results in the spherically-symmetric case}

\begin{figure}
\includegraphics[height = 0.4\textheight, width
=1\textwidth]{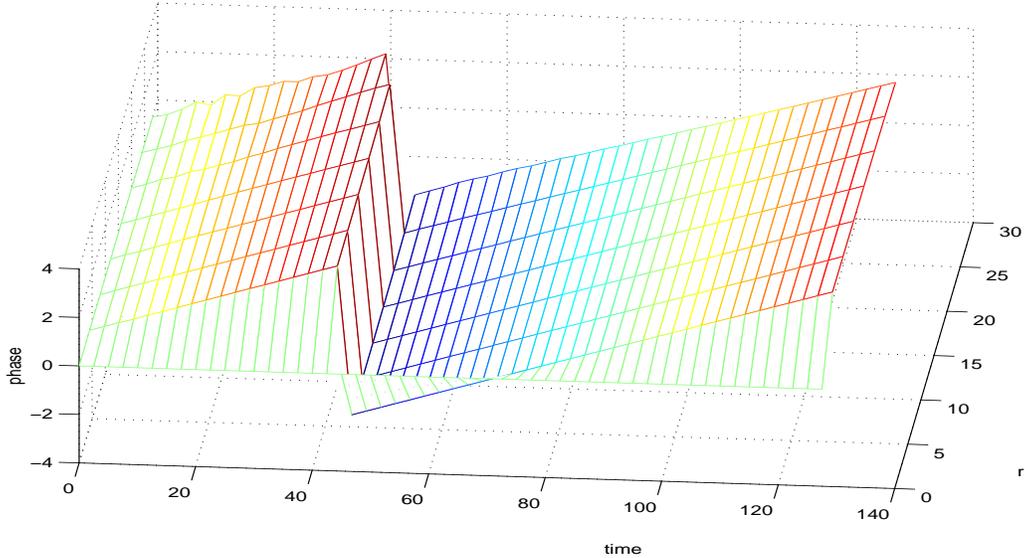}
\caption{The phase angle of the ground state as a function of time.}
\label{cphase}
\end{figure}

Having tested the method and the sponges satisfactorily, we 
evolve the ground state, that is we use as initial data
the stationary-state of lowest energy which is known from \cite{cq98}
or \cite{hmt1}. Since this is a stationary state it should evolve with
a factor $e^{-iEt}$ and this is just what we find (see
figure~\ref{cphase}: the phase is linear in time).  
However if this evolution is continued for a long enough time, numerical errors
accumulate and the solution drifts away from the ground state. By
Fourier 
analysing the solution one finds lines in the power spectrum at
(approximately) the
normal frequencies for a perturbation about the ground state already 
found by linear perturbation theory in \cite{hmt1}. Thus we really are
getting the stable but perturbed ground state. We have also evolved
from data which are the ground state with a small perturbation
introduced by hand and the picture is the same.

\begin{figure}
\includegraphics[height = 0.4\textheight, width =
1\textwidth]{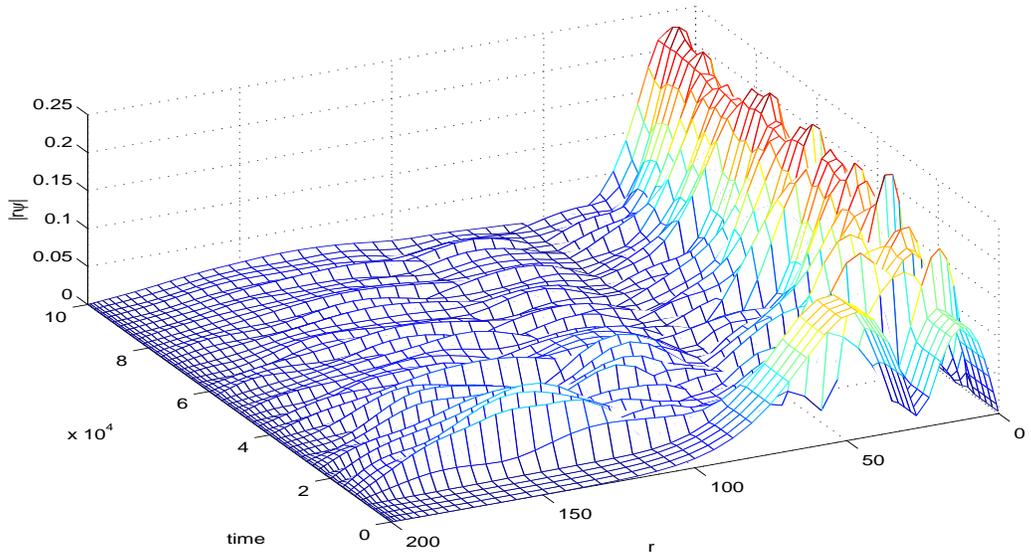}
\caption{Evolution of second state.}
\label{num2p01}
\end{figure}

Next we evolve the second spherically-symmetric stationary state taken 
from \cite{hmt1}. We should expect this also to evolve as a stationary
state at least for a while until numerical errors accumulate. What we
obtain is shown in figure~\ref{num2p01}. This shows $|r\psi|$, so that
the second state, which has a zero, appears as bimodal. The evolution
proceeds as a stationary-state to about $t=2000$ and then there is a
sudden change to a unimodal solution with some oscillation. This we
claim is a perturbed and rescaled ground state solution, 
with total probability less than one.

\begin{figure}
\includegraphics[height = 0.4\textheight, width =
1\textwidth]{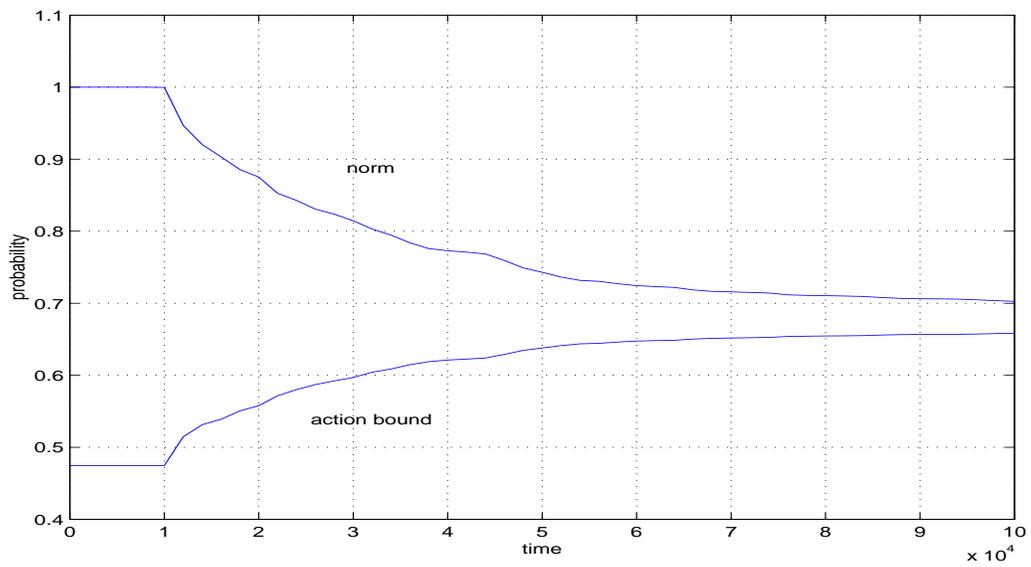}
\caption{Decay of the second state: the two measures of probability converging}
\label{num2p04}
\end{figure}

To support this claim, we have again calculated the power spectrum and
find lines at (approximately) the normal frequencies of perturbations about the ground
state (to find agreement we need to rescale these frequencies with the
factor corresponding to the rescaling of the residual ground state). 
We can also use the check suggested by
equation~(\ref{bound}). That is, we calculate both the probability $p$ and
the action or conserved energy ${\mathcal{E}}$
remaining on the grid at time $t$ and compute the bound on $p$
provided by (\ref{bound}). If the argument leading to this bound is
correct then these two probabilities should converge on each other and
as we see in figure~\ref{num2p04} they do.
Again, we can evolve from the second state with a small perurbation
introduced by hand rather than waiting for the numerical errors to
accumulate when the collapse to the ground state is more rapid, or
alternatively we can evolve from the second state but with a more
stringent tolerance on the iteration determining $\phi$, when the
collapse takes longer.

\begin{figure}
\includegraphics[height = 0.4\textheight, width =
1\textwidth]{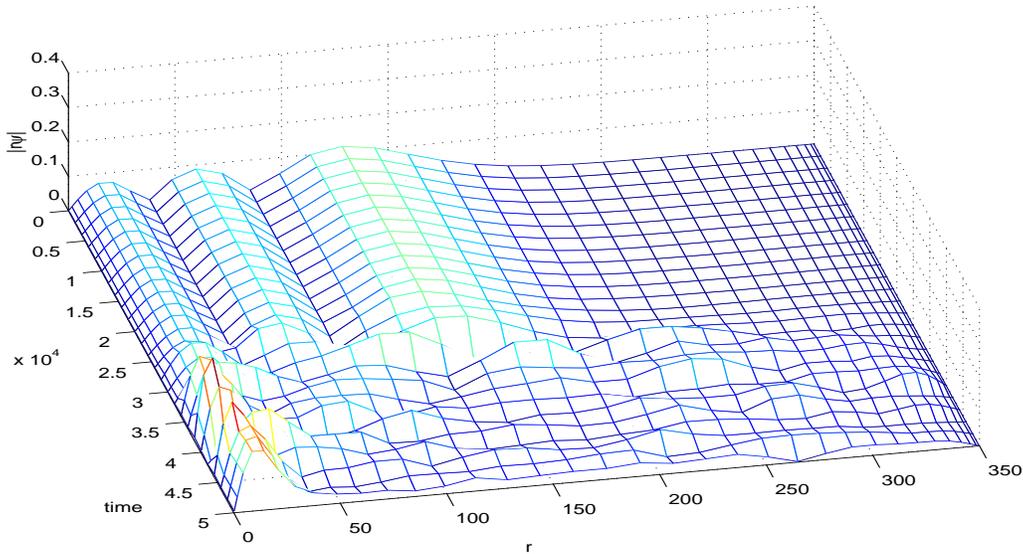} 
\caption{Evolution of third state.}
\label{sne3}
\end{figure}

Moving on, the decay of the third state is shown in figure~\ref{sne3}. The
initial state this time has three maxima and evolves for a while as a
stationary state before collapsing to a rescaled ground state with the
emission of some probability to infinity.
We have plotted the figure corresponding to figure~\ref{num2p04} and
we again find convergence.\\

The picture that emerges from these calculations is of nonlinear 
instability of all states after
the ground state. Each higher state decays to the ground state either
by the accumulation of numerical errors or because of an explicitly
included perturbation. The end result is a noisy rescaled ground
state, noisy because the eigenvalues for perturbation about the ground
state are all imaginary.\\

\begin{figure}
\includegraphics[height = 0.4\textheight, width =
1\textwidth]{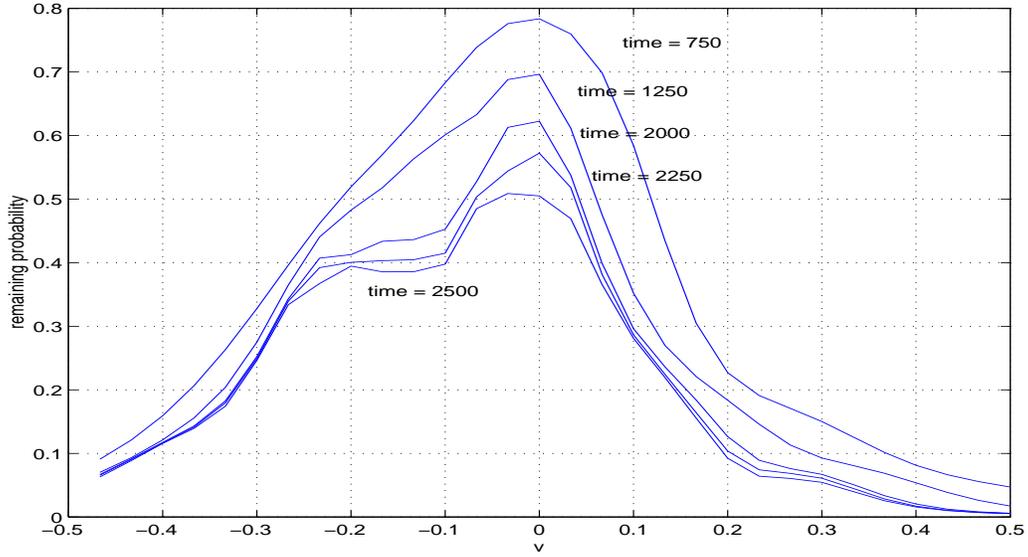} 
\caption{Evolution of Gaussians with varying $v$: $\sigma=6$, $a=50$.}
\label{ass6p2}
\end{figure}

\begin{figure}
\includegraphics[height = 0.4\textheight,width =
1\textwidth]{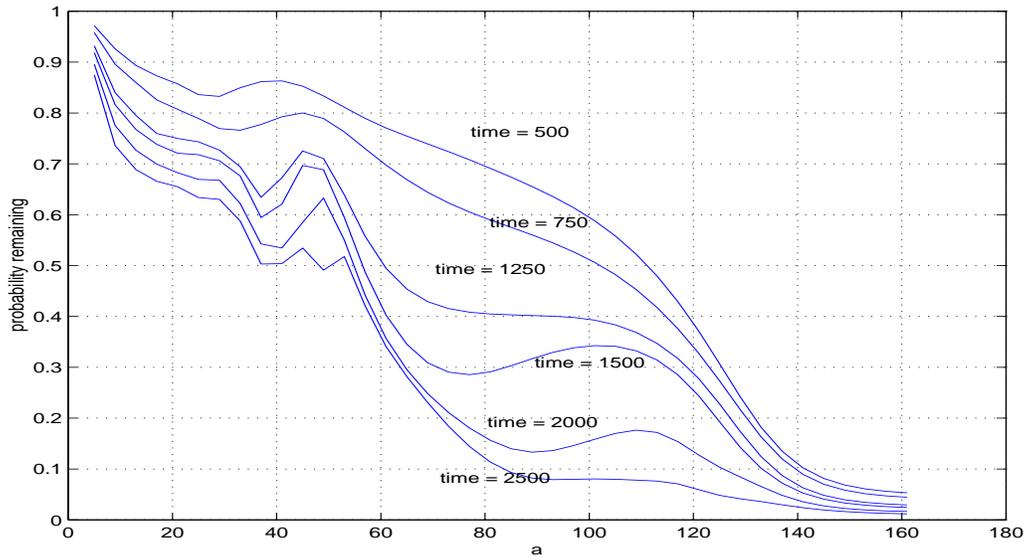}
\caption{Evolution of Gaussians with varying $a$: $\sigma=6$, $v=0$.}
\label{ass6p5}
\end{figure}

\begin{figure}
\includegraphics[height = 0.4\textheight,width = 1\textwidth]{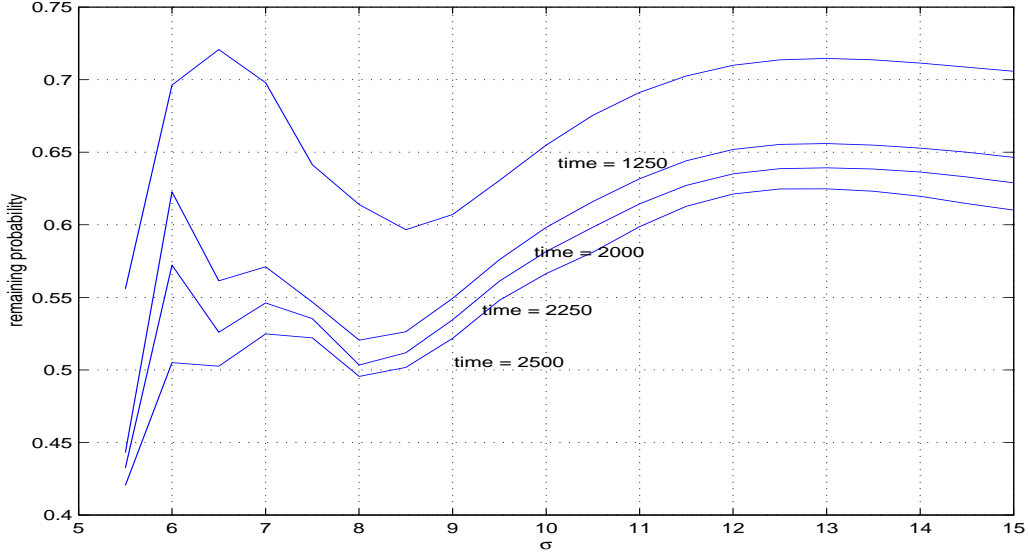}
\caption{Evolution of Gaussians with varying $\sigma$: $v=0$, $a=50$.}
\label{ass6p9}
\end{figure}

The other spherically symmetric evolutions which we have calculated
are with the initial data furnished by equation~(\ref{lexp}) with
$t=0$. This is a Gaussian bump centered at $r=a$ with width $\sigma$
and moving with velocity $v$. The conserved energy of this data rises
with $v$ (because the kinetic energy rises) and with $\sigma$ (because
the particle is more localised). In all cases the solution disperses
leaving a rescaled ground state which we can characterise by the
probability residing in it. In figures~\ref{ass6p2}, \ref{ass6p5}
and \ref{ass6p9} we plot the residual probability against $v$, $a$ 
and $\sigma$ at different times.

From figure \ref{ass6p2} we see that the residual probability is
greatest if the lump is released from rest, but it is easier to trap 
probability if the initial velocity is ingoing than if it is outgoing. From
figure~\ref{ass6p5} we see that for lumps released from rest, the
residual probability is greater if the lump starts closer in, or in
other words is more gravitationally bound. From figure~\ref{ass6p9}
we see that reducing $\sigma$, which raises the energy, leads to more 
dispersion.

A key motivation for this set of calculations was the desire to check
convergence of the method, and we have done this in a
variety of ways. First for the time-step, with a Crank-Nicholson
method we expect to
have quadratic convergence, and this can be
investigated with a Richardson quotient. We suppose that for some
variable of interest the calculated value $O_h$ and actual values $I$
are related by
\[O_h \sim I + Ah^k, \]
where the time step size is $h$, and $k$ is the order of the error, 
then we can calculate the Richardson quotient:
\[ \frac{O_{h_1} - O_{h_3}}{O_{h_2}-O_{h_3}}, \]
where $h_1$,$h_2$ and $h_3$ are three different values of the step
size. With $k$ known, this will be a simple function of the $h_i$ and
so will provide a check on $k$. Next for convergence in space we can repeat the
calculation with different numbers $N$ of Chebyshev points. For both
checks, the
results converge as required.\\

With this calculation we confirm the picture of the SN evolution which
we have been claiming: the Schr\"odinger equation tends to disperse
probability but the Newtonian gravitational attraction holds it
together; all states after the ground state are unstable and the
evolution in general leads to a dispersion of probability to 
infinity, leaving some residual
probability in a rescaled ground state.

\section{The axially-symmetric SN equations}

 In this chapter, we solve the SN equations
for an axially-symmetric system in 3 dimensions. The wave-function is now
a function of polar coordinates $r$ and $\theta$ but is independent of
polar angle. We shall first find stationary
solutions. These typically have nonzero total
angular momentum and include a dipole-like solution which appears to
minimise the energy  among wave functions which are odd in (the usual) $z$. We
then consider the time-dependent problem. In particular we evolve the
dipole-like state, and it turns out to be nonlinearly unstable - the
two regions of probability density attract each other and fall
together leaving a multiple of the ground
state as the evolutions in Section 2 did.
  
With $u = r\psi$ the system of equations~(\ref{ndimS}) becomes:
\begin{eqnarray}
iu_t &=& -u_{rr} - \frac{1}{r^2\sin\theta}(\sin\theta{}u_{\theta})_{\theta}
+\phi{}u,\label{axiSN1} \\
\frac{|u|^2}{r} &=& (r\phi)_{rr} + \frac{1}{r\sin\theta}
(\sin\theta{}\phi_{\theta})_{\theta}.\label{axiSN2}  
\end{eqnarray}
For stationary solutions, the left-hand-side of (\ref{axiSN1}) is
replaced by $Eu$. Boundary conditions are that $u=0$ at $r=0$ and
$r=L$, $\phi=0$
at $r=L$ and finite at $r=0$, and $u_{\theta} = 0 =\phi_{\theta}$ 
at $\theta = 0$ and $\theta = \pi.$

To find stationary solutions we proceed as follows:

\begin{enumerate}
\item take as an initial guess for the potential $\displaystyle{\phi =
\frac{-1}{(1+r)}}$;  
\item using this potential solve the time-independent
Schr\"odinger equation;    
\item select an eigenfunction in such a way that the procedure will
converge to a stationary state (this needs trial and error); 
\item calculate the
potential due to the chosen eigenfunction;    
\label{newpot}
\item take the new potential to be the one obtained from step~\ref{newpot}
above but symmetrised, since we require that $\phi$ should be
symmetric around the $\displaystyle{\theta = \frac{\pi}{2}}$
(otherwise numerical errors can cause 
the wave-function to move along the axis);
\item now provided that the new potential does not differ from the 
	previous potential by some fixed tolerance in the norm 
	of the difference, stop, otherwise continue from step 2. 
\end{enumerate}

In step 2 we solve the eigenvalue
problem by a 2-dimensional spectral method, using Chebyshev 
differentiation in the directions of $r$ and $\theta$. In step 3, the
eigenfunctions at the first iteration are labelled by the usual $l, m$
and $n$ quantum numbers, though with $m=0$ for axisymmetry. The idea
is to choose one and run through the iteration, hoping for
convergence. When the method converges, we do obtain a stationary
state but we do not arrive at a one-to-one correspondence between
solutions of the starting linear problem and the final nonlinear problem.

For the energy eigenvalue $E$ (which is not the conserved energy) we
have the formula
\begin{equation}
E = \int |\nabla\psi|^2 + \phi|\psi|^2 ,
\end{equation}
while the total angular momentum $J^2$, after integration by parts, is :
\begin{equation}
J^2 = \int|\frac{\partial\psi}{\partial\theta}|^2
\end{equation}
Both integrals are over ${\bf R}^3$.

We present in Table~\ref{sae7} the first few stationary states of
the axially symmetric solution of the SN equations ordered by their
energy and named for convenience axi1 to
axi8.

\begin{table}[ht] 
\begin{center}
\begin{tabular}{|r|r|c|}
\hline
Energy & $J^2$ & name \\  
\hline
-0.1592 & zero  & axi1 \\ 
-0.0599 & 5.1853 & axi2 \\
-0.0358 & 0.002  & axi3 \\
-0.0292 & 2.3548 & axi4 \\
-0.0263 & 17.155 & axi5 \\%formerly axp7
-0.0208 & 3.1178 & axi6 \\%formerly axp8
-0.0162 & 5.2053 & axi7 \\%formerly axp5
-0.0115 & 1.9E-6 & axi8 \\%formerly axp6
\hline 
\end{tabular}
\end{center}
\caption{The first few axially-symmetric stationary states.}
\label{sae7}
\end{table}

\begin{figure}
\includegraphics[height = 0.4\textheight, width =
1\textwidth]{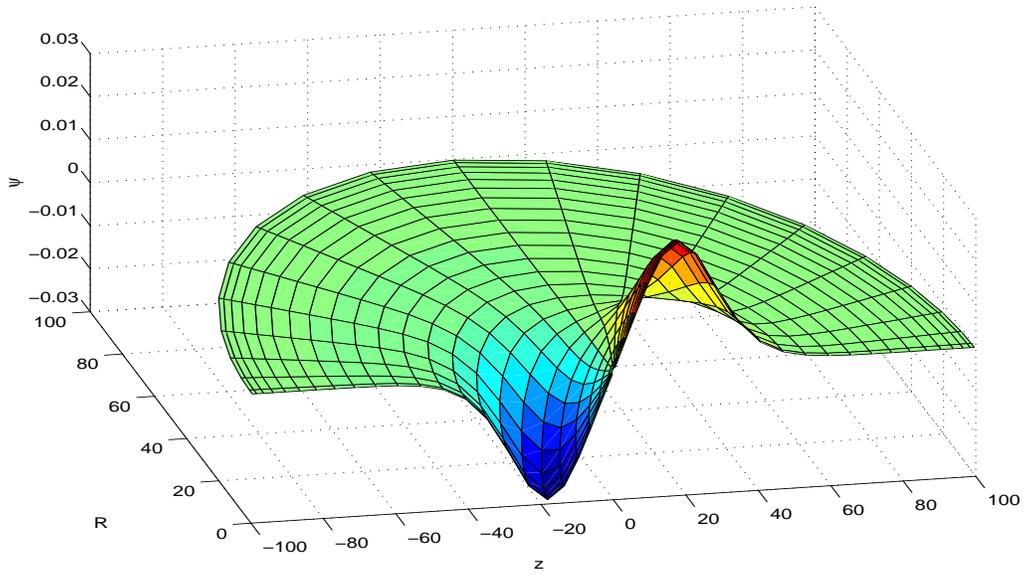}
\caption{The dipole-like state, axi2.}
\label{axp2}
\end{figure}

\begin{figure} 
\includegraphics[height = 0.4\textheight, width =
1\textwidth]{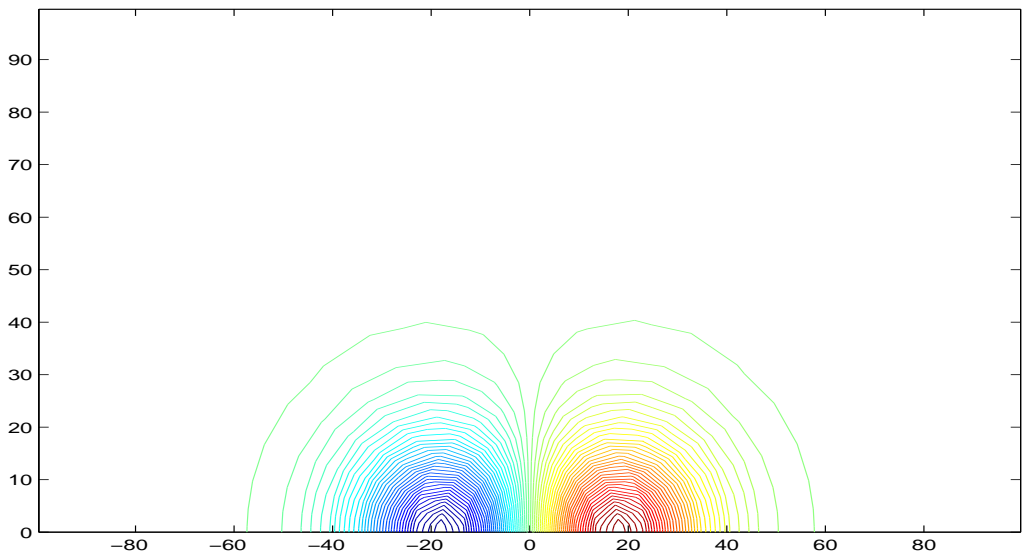} 
\caption{Contour Plot of the dipole, axi2.}
\label{axp2c}
\end{figure}

\begin{figure}
\includegraphics[height = 0.4\textheight, width =
1\textwidth]{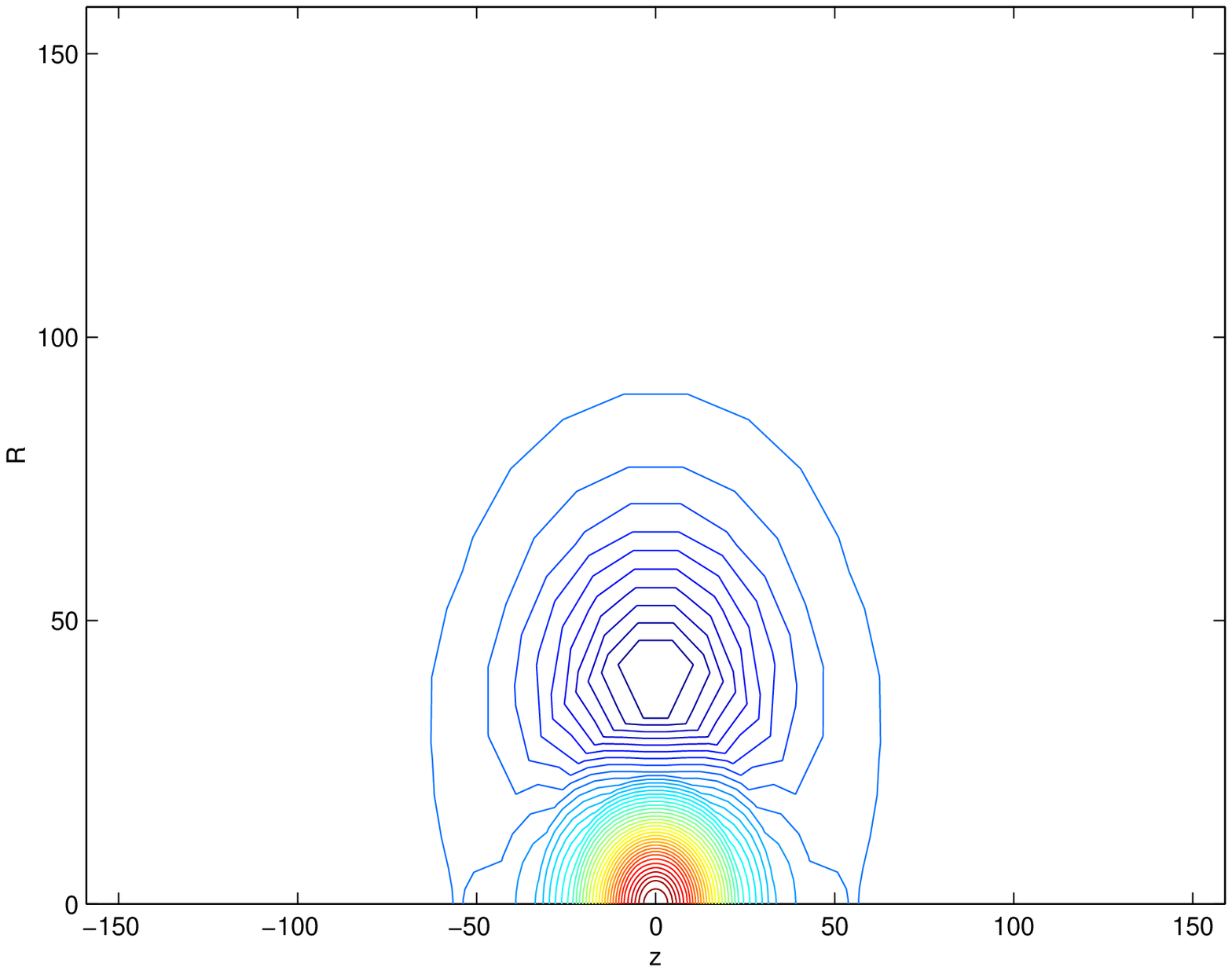}
\caption{ Contour plot of the state axi4.}
\label{axp5c}
\end{figure}

\begin{figure}
\includegraphics[height = 0.4\textheight, width =
1\textwidth]{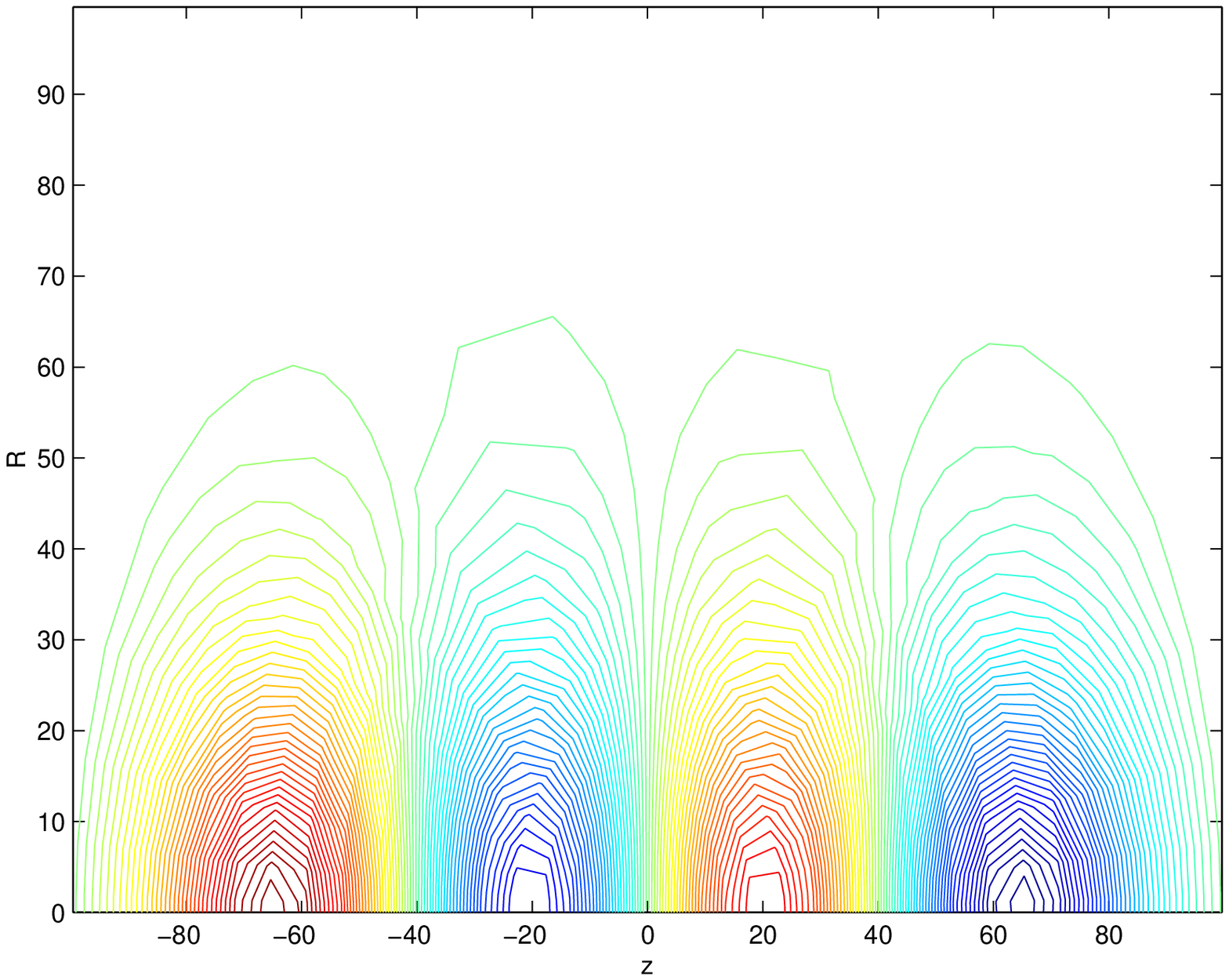}
\caption{Contour plot of the state axi5.}
\label{axp8c}
\end{figure}

\begin{figure}
\includegraphics[height =0.4\textheight, width =
1\textwidth]{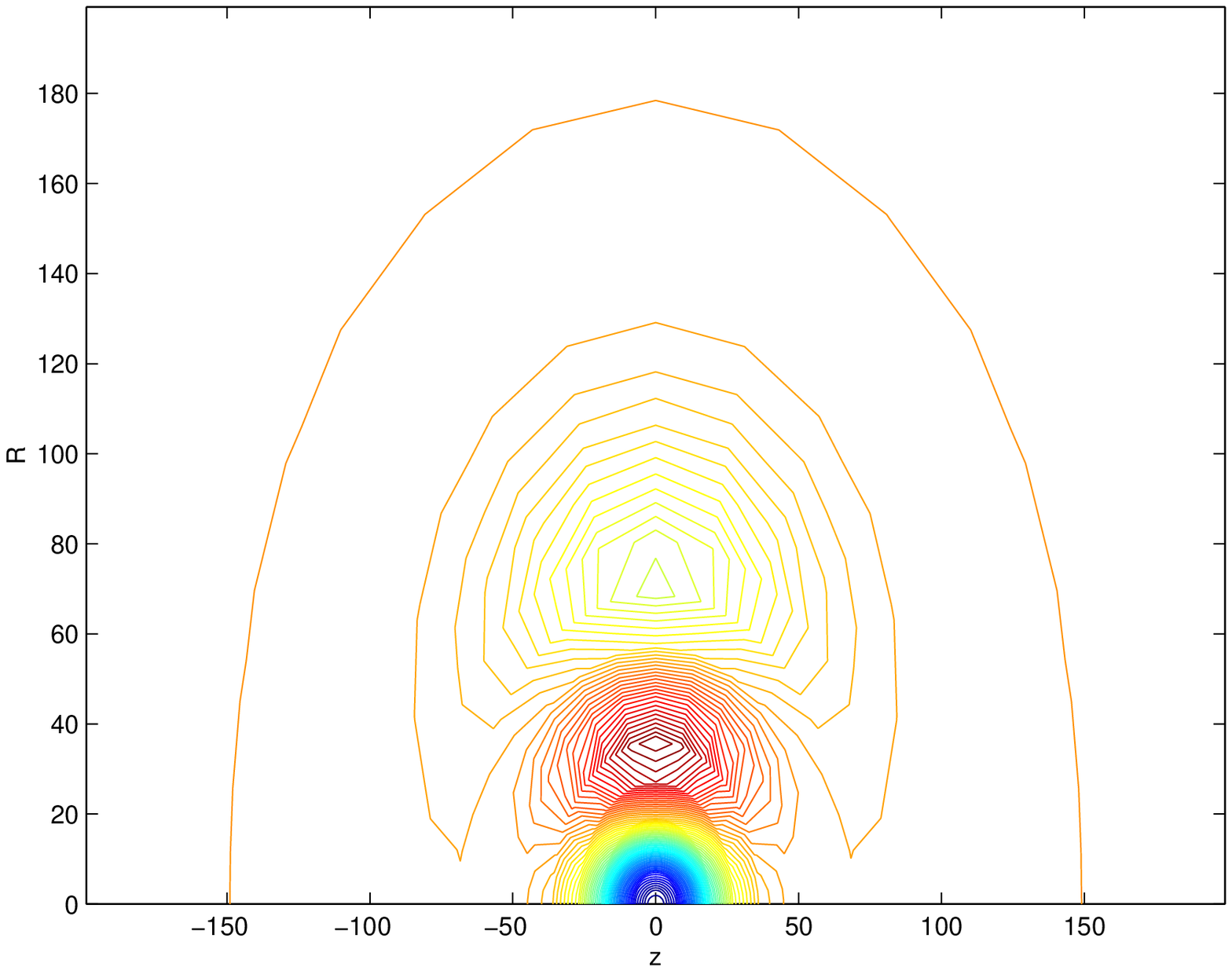}
\caption{Contour plot of the state axi6.}
\label{axp9c}
\end{figure}

\begin{figure}
\includegraphics[height = 0.4\textheight, width =
1\textwidth]{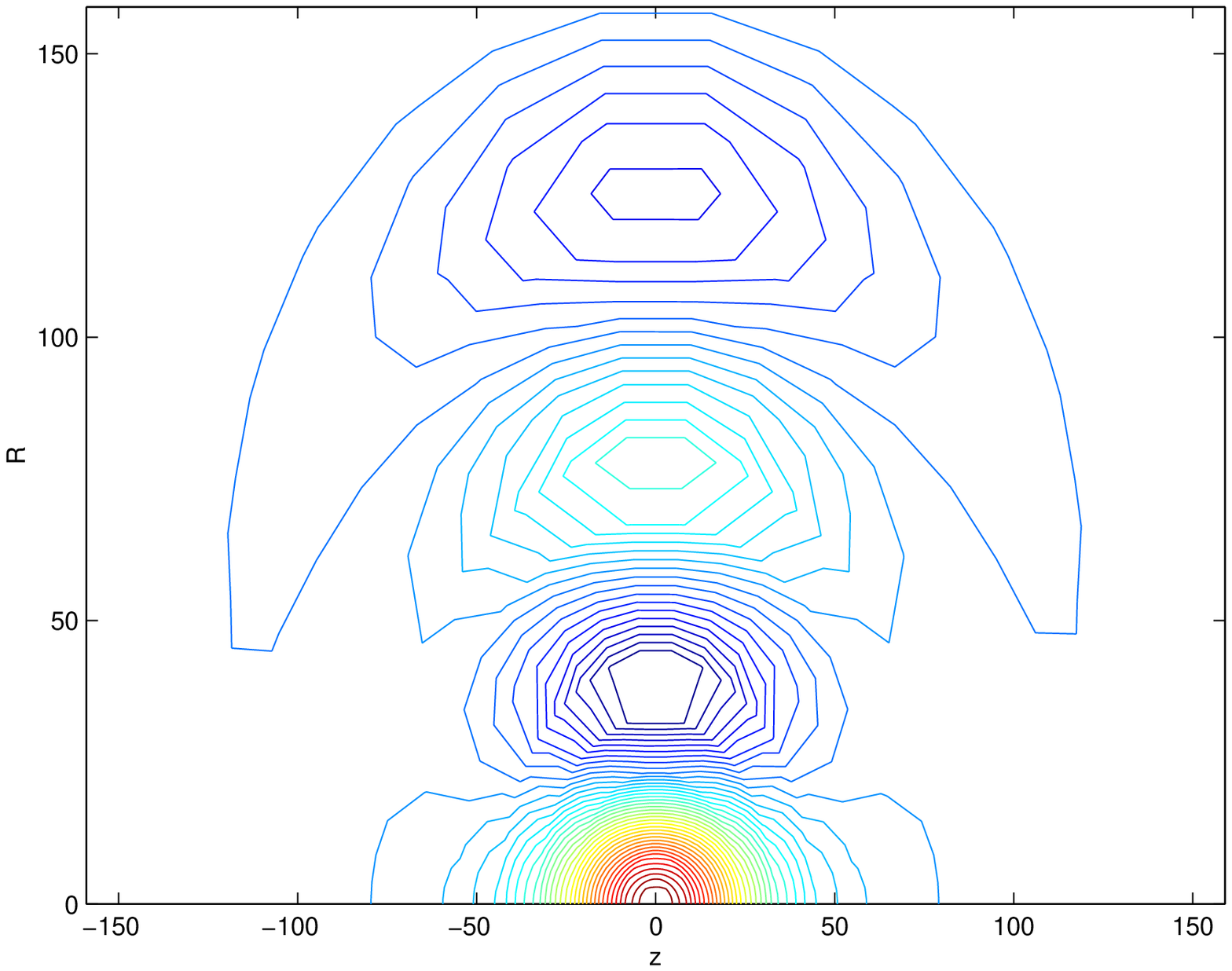}
\caption{Contour plot of the state axi7.}
\label{axp6c}
\end{figure}

In this table, axi1, axi3 and axi8 are spherically-symmetric solutions
turning up again (as they should). Axi2, shown as a surface 
in figure~\ref{axp2} and as a contour plot in figure~\ref{axp2c}, is
very much like a dipole solution and appears to be the solution
minimising the energy among wavefunctions odd in $z$. If so it was
found already in \cite{sb95}. Contour plots of axi4-axi7 are given in 
figures~\ref{axp5c} to \ref{axp6c}.

To solve the time-dependent axisymmetric SN-equations, we shall use an
alternating direction implicit (or ADI) method 
(see e.g.\cite{f3}, \cite{Gthesis} or \cite{mm}) We split the Laplacian
in (\ref{axiSN1}) to write it as
\begin{equation}
\dot{u}=i(L_1+L_2-\phi)u
\end{equation}
where
\begin{eqnarray}
L_1 & = & \frac{1}{r}\frac{\partial^2}{\partial{}r^2}\\ 
L_2 & = & \frac{1}{r}[\frac{\partial^2}{\partial{}\theta^2} +
\cot\theta\frac{\partial}{\partial\theta}].
\end{eqnarray}
Next we introduce new variables $S$ and $T$ and write this in the
formally equivalent form:
\begin{eqnarray}
\exp(-\frac{ih}{2}L_1)S(t) & = & \exp(\frac{ih}{2}L_2)u(t)\nonumber\\
\exp(-\frac{ih}{2}L_2)T(t) & = & \exp(\frac{ih}{2}L_1)S(t)\label{ADI1}\\
\exp(\frac{ih}{2}\phi)u(t+h) & = & \exp(-\frac{ih}{2}\phi)T(t)\nonumber
\end{eqnarray}
where we suppress the dependence on $r$ and $\theta$. To obtain a discrete
form of (\ref{ADI1}) we linearise in the time-step $h$ to find
\begin{eqnarray}
(1-\frac{ih}{2}L_1)S^{n} & = & (1+\frac{ih}{2}L_2)u^n\nonumber\\
(1-\frac{ih}{2}L_2)T^{n} & = & (1+\frac{ih}{2}L_1)S^n\label{ADI2}\\
(1+\frac{ih}{2}\phi^{n+1})u^{n+1} & = & (1-\frac{ih}{2}\phi^n)T^n\nonumber
\end{eqnarray}
where the superscript indicates the value of discretised time.

For the Poisson equation (\ref{axiSN2}) we have
\[
(L_1+L_2)(r\phi)=\frac{1}{r^2}|u|^2
\]
which we solve at each instant by a Peaceman-Rachford ADI 
iteration \cite{f3}. This means we discretise it as
\begin{eqnarray}
(L_1+\rho)\phi_{k+1} & = & -(L_2-\rho)\phi_k+\frac{1}{r^2}|u|^2\label{ADI3}\\
(L_2+\rho)\phi_{k+1} & = & -(L_1-\rho)\phi_k+\frac{1}{r^2}|u|^2\nonumber
\end{eqnarray}
and iterate with $\phi^0=0$. Here $k$ labels the iteration 
(at a fixed time) and $\rho$ is a small
constant chosen so that the iteration converges.

The boundary conditions are that $u$ and $\phi$ vanish at the outer
boundary and so we need sponges as in Section 2 to prevent waves of
probability reflecting back.

\begin{figure}
\includegraphics[height = 0.4\textheight, width =
1\textwidth]{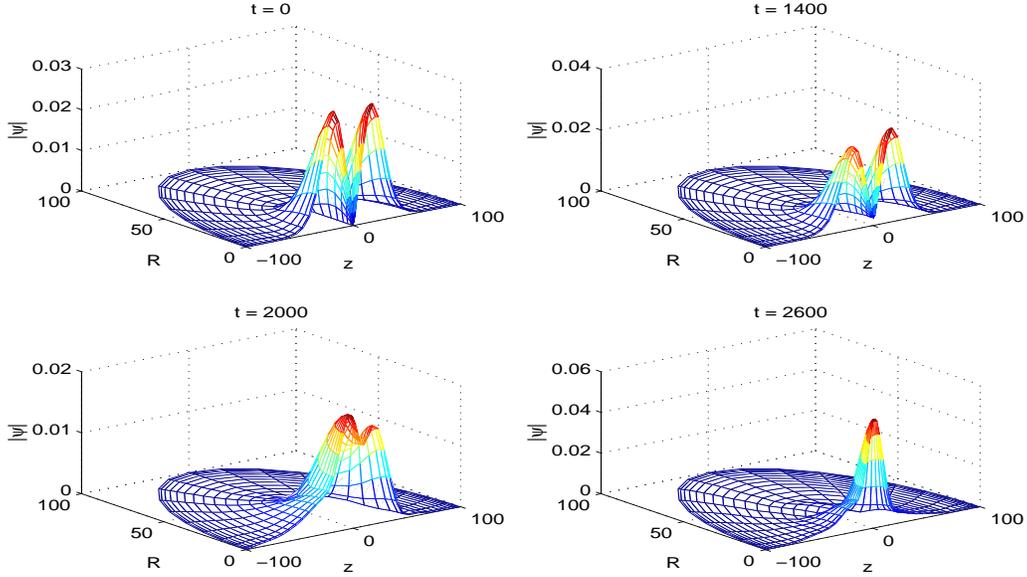} 
\caption{Evolution of the dipole.}
\label{die}
\end{figure}

As an example, we take as initial data the dipole state of 
figure~\ref{axp2}. The evolution of $|\psi|$ for this is shown in 
figure~\ref{die}. (Note that $\psi$ is initially real but must
become complex before again becoming approximately real in the remote
future. The initial data for $\psi$ is odd as a function of $z$ but it
ends up approximately even.)

Following the lessons learned in Section 3, we should 
expect the stationary state to be unstable. We can compute the 
evolution for short
times to see that initially it remains a stationary state, that is
that the only change is a phase growing linearly with time. However 
when numerical errors have built up the state becomes unstable and the two
concentrations of probability density fall into each
other. Probability leaves the grid, as does angular momentum, and we
are left with a multiple of the ground state.
We can check convergence of the method by calculating a Richardson
quotient with three different time-steps (to find that convergence is
now linear in time), and by varying the number of
Chebyshev points.

What we learn from this calculation is that, as well as dispersion
which is what we mostly saw in Section 3, the solutions of the SN
equation show gravitational attraction: lumps of probability density
released from rest fall into each other. In the next section we shall
see that, at least for the two-dimensional SN equations, lumps of
probability can orbit each other.

\section{The two-Dimensional SN equations}

In this section, we shall consider the SN equations in a plane, that is
in Cartesian coordinates $x$,$y$. We shall find a dipole-like stationary
solution, and some solutions which are like rigidly rotating
dipoles. These rigidly rotating solutions are unstable however and
will merge, radiating angular momentum.

The SN equations in this case are
\begin{eqnarray} 
i\psi_t & = & -\psi_{xx}-\psi_{yy} + \phi\psi \\
\phi_{xx}+ \phi_{yy}& = & |\psi|^2\nonumber
\end{eqnarray}
We use the ADI scheme as in equation (\ref{ADI2}) but with 
the understanding that
now
\begin{eqnarray*}
L_1 & = & \frac{\partial^2}{\partial x^2}\\
L_2 & = & \frac{\partial^2}{\partial y^2}.
\end{eqnarray*}
For the potential we use the counterpart of 
(\ref{ADI3}) with the same understanding. The
boundary conditions are that $\psi$ and $\phi$ vanish at the edges of
the grid, which is now a large square. We still need sponges and we do
not want them to have corners so we take functions like 
$s(x,y) = \min[1,e^{0.5(\sqrt(x^2+y^2)-20)}]$. We can test the efficacy of
the sponges by evolving moving two-dimensional Gaussians to see that
they propagate off the grid, as they do.

Once we have the code, we can look for solutions corresponding to
those found already in 3-dimensions. In particular we find a stable
ground state and a
dipole-like solution which is stable for a while before decaying to
the ground state. We can seek solutions with no counterpart among those found
already by making an ansatz of rigid rotation. That is we look for
solutions which in polar coordinates $r,\theta$ take the form:
\begin{equation}
\psi(r,\theta,t) = e^{-i{}Et}\psi(r,\theta+\omega{}t)
\label{rr}
\end{equation}
for (real) constants $E$ and $\omega$. (Solutions of a similar nature
in 3-dimensions would depend on all three spatial coordinates which is
why we haven't seen them so far.)

\begin{figure}
\includegraphics[height = 0.4\textheight, width =
1\textwidth]{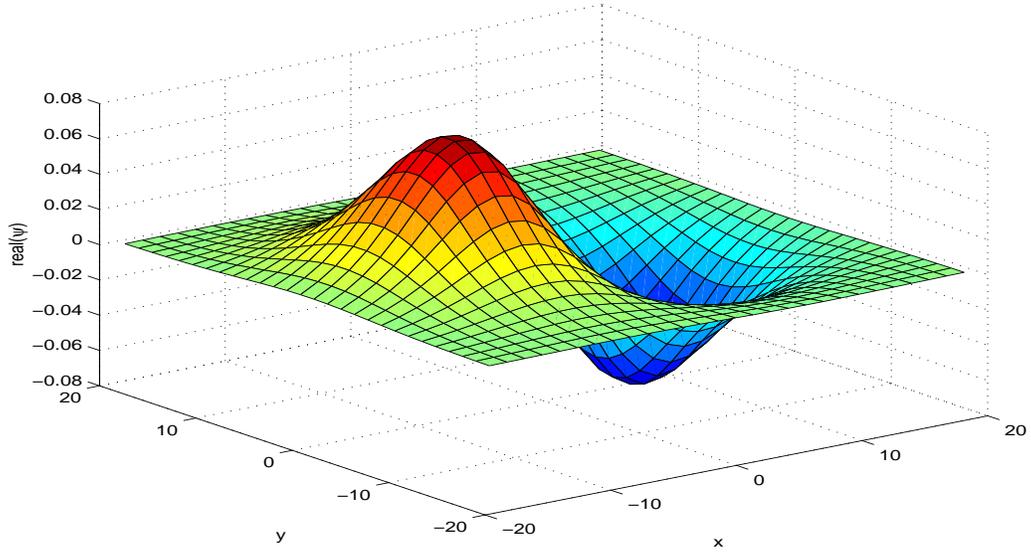}
\caption{Real part of spinning solution, $\omega=0.005$.}
\label{ass8sp1}
\end{figure}

\begin{figure}
\includegraphics[height = 0.4\textheight, width =
1\textwidth]{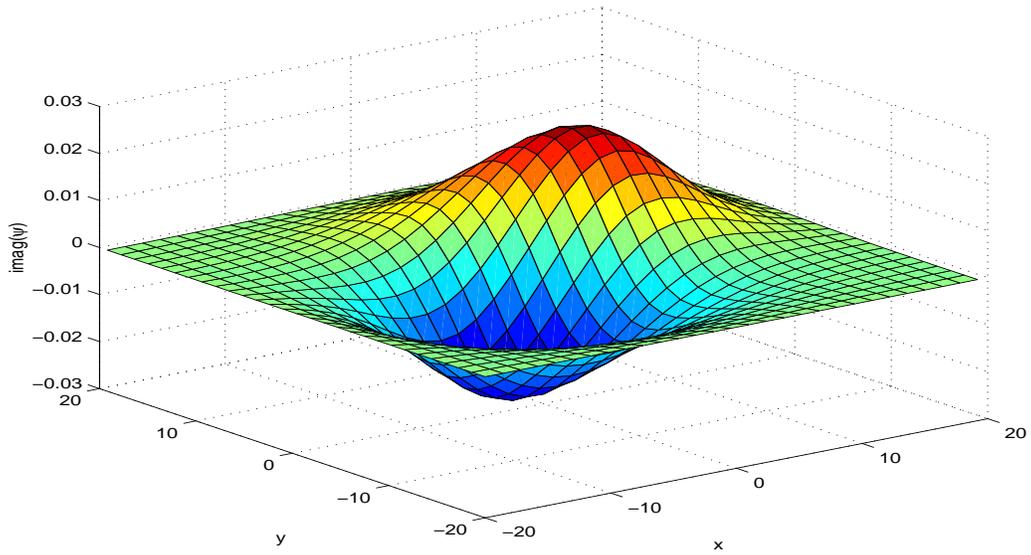}
\caption{Imaginary part of spinning solution, $\omega=0.005$.}
\label{ass8sp2}
\end{figure}

To separate the SN equations
for a solution like (\ref{rr}), we go into the rotating frame with Cartesian
coordinates $X,Y$ given by
\begin{eqnarray*}
X = x\cos\omega{}t+y\sin\omega{}t, \\
Y = -x\sin\omega{}t+y\cos\omega{}t,
\end{eqnarray*}
and the time-dependence separates off to leave the equations as
\begin{eqnarray}
-\nabla^2\psi + \phi\psi- i\omega{}(Y\psi_X -X\psi_Y) & = &E\psi\label{rot1}\\
\nabla^2\phi &=& |\psi|^2,\nonumber
\end{eqnarray}
To solve (\ref{rot1}), we begin with a small value of $\omega$ and the
dipole-like solution. Iteration leads to a solution, and we can study
its change with increasing $\omega$. The wave-function $\psi$ is
necessarily complex and we display the real and imaginary parts of 
such a solution, with 
$\omega =0.005$, in figures~\ref{ass8sp1} and~\ref{ass8sp2}.

\begin{figure}
\includegraphics[height = 0.4\textheight, width =
1\textwidth]{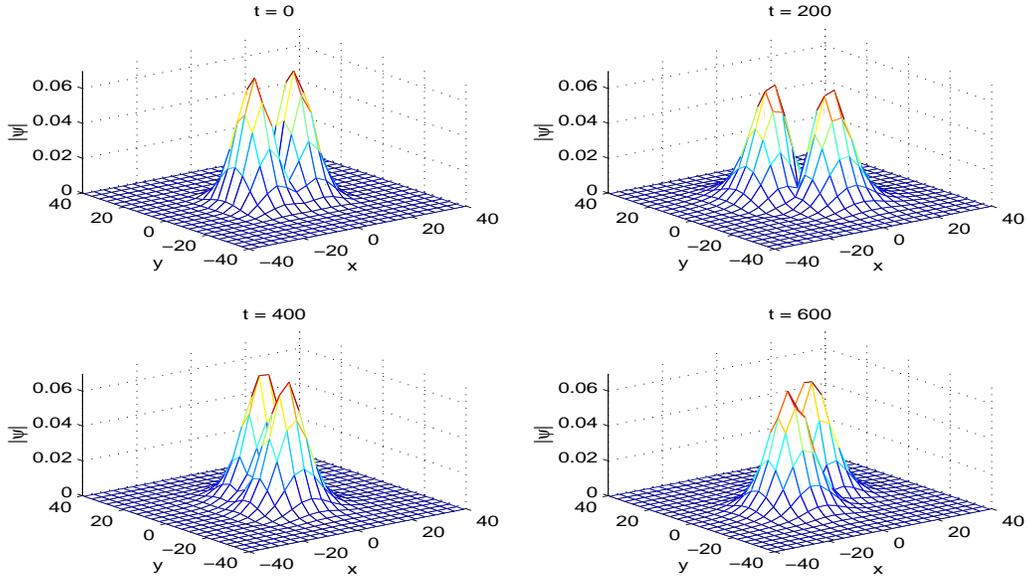}
\caption{Initial evolution of spinning solution: orbiting.}
\label{ass8sp3}
\end{figure}

\begin{figure}
\includegraphics[height = 0.4\textheight, width =
1\textwidth]{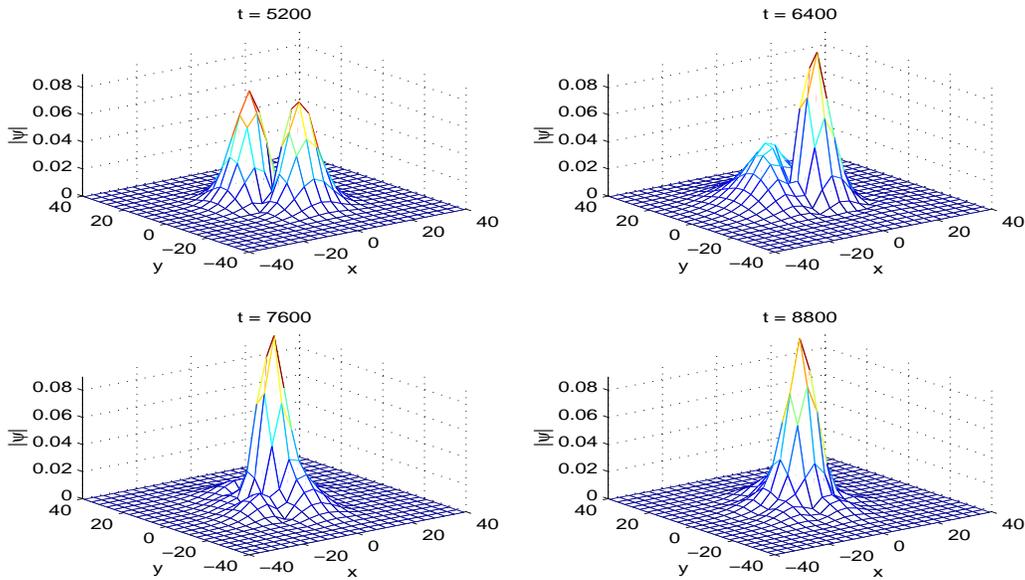}
\caption{Later evolution of spinning solution: collapse.}
\label{ass8sp4}
\end{figure}

The solution just found is a stationary state but we must expect it to
be unstable. If we use it as initial data for the time-dependent
problem then we find as before that it evolves as a stationary state
for some time (see figure~\ref{ass8sp3}) before numerical errors build
up and it becomes unstable (figure~\ref{ass8sp4}). The two lumps of
probability orbit each other about four times before collapsing into a
single lump in about one orbital period. Probability and angular
momentum are radiated off of the grid.

What we have found by this calculation is a confirmation of the
picture of lumps of probability interacting gravitationally with each 
other. Here the
lumps are in orbit around each other for a while before becoming
unstable and collapsing into a single lump. By earlier work, we must
then expect the
probability to disperse leaving a rescaled ground state, as it does.

\section*{Acknowledgement}
The work described in this paper formed part of the D.Phil. thesis of
the first author and he gratefully acknowledges the receipt of a grant
from EPSRC.

\end{document}